# PEDESTRIAN FLOW AT BOTTLENECKS – VALIDATION AND CALIBRATION OF VISSIM'S SOCIAL FORCE MODEL OF PEDESTRIAN TRAFFIC AND ITS EMPIRICAL FOUNDATIONS

*Tobias Kretz, Stefan Hengst, and Peter Vortisch*
*PTV AG*
*Stumpfstraße 1*
*D-76131 Karlsruhe, Germany*
`{Tobias.Kretz, Stefan.Hengst, Peter.Vortisch}@PTV.De`

## ABSTRACT

In this contribution first results of experiments on pedestrian flow through bottlenecks are presented and then compared to simulation results obtained with the Social Force Model in the Vissim simulation framework. Concerning the experiments it is argued that the basic dependence between flow and bottleneck width is not a step function but that it is linear and modified by the effect of a psychological phenomenon. The simulation results as well show a linear dependence and the parameters can be calibrated such that the absolute values for flow and time fit to range of experimental results.

## INTRODUCTION

Microscopic models of the dynamics of pedestrians can be validated and calibrated with respect to at least two concerns: the reproduction of the interesting macroscopic observables and a natural microscopic "look and feel" of the individual agents. Not only can the latter one be achieved without the former one, but also vice versa, the former one without the latter one, when parts of the microscopic behavior are not relevant for the macroscopic observables or even when two kinds of relevant unrealistic microscopic effects cancel mutually.

Concerning the validation of the resulting macroscopic observables, there are typically two scenarios which are used for quantitative validation: the flow through bottlenecks and the fundamental diagram. This contribution is about the flow through bottlenecks.

For the flow through bottlenecks it has for almost one hundred years [1, 2] (see figure 1) been discussed whether there is a stepwise increase in dependence of the width of the bottleneck – this was related to the idea of a certain number of lanes that fit into the bottleneck – or a linear dependence. While there have always been reports, that the flow increases linearly with the width, the "stepwise school" has long been dominant. However, recent experiments have been con-ducted that clearly support the linear dependence hypothesis [3–7]. The results of these experiments will be compared to simulations of the Social Force Model of Helbing and Molnar [8, 9] in the Vissim framework.

It is not only the flow through the bottleneck, but also oscillations in counter-flow situations at bottlenecks [10] and the shape of the waiting crowd in front of the bottleneck [11, 12] that is suitable for validation and comparison with reality, as this shape can be near one-dimensional (a queue of people which is just as wide as the bottleneck itself), two-dimensional (a half circle) or anything in between (a drop-like shape), but due to limited space, such considerations are left out here.

## RESULTS OF BOTTLENECK EXPERIMENTS

As figure 2 shows a multitude of recent experiments gave results that are in favor of a more or less linear dependence of the flow with the bottleneck width. Out of the experiments from figure 2 those have been chosen for this contribution to be compared to simulation results, which have at least three measurements at or below 1 m [3–7]. Figure 3 shows an overview of these results. For the total time the results have been scaled such as if 100 persons had participated in the experiment, accepting possible errors as including the time between the start signal and the moment when the first pedestrian leaves into the scaling, as the main focus is not on absolute values but on the overall dependence of the flux on the bottleneck width. Therefore it has been assumed that the total time for N pedestrians consists of N-1 time gaps. The scaling factor therefore was 99/(n-1), if n pedestrians participated. (From [6] the results of the experiments with 80 participants were used.) Except for the scaling process total time in experiments and simulation is the time between the start signal and the time when the last pedestrian passes the bottleneck. The flux is calculated from this time and the number of pedestrians (implying that for one pedestrian no flux would be defined). Specific flux is flux over bottleneck width.

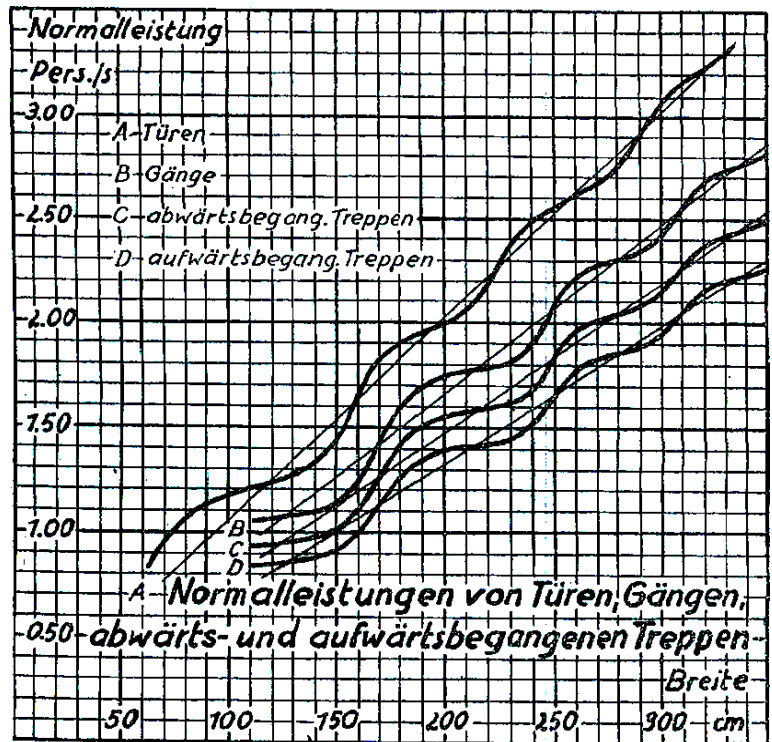

**Figure 1:** Snake line shaped dependence of the flow in dependence of the bottleneck width as published in [2]. The snake line is explained to form as a linear dependence modulated by a step effect.

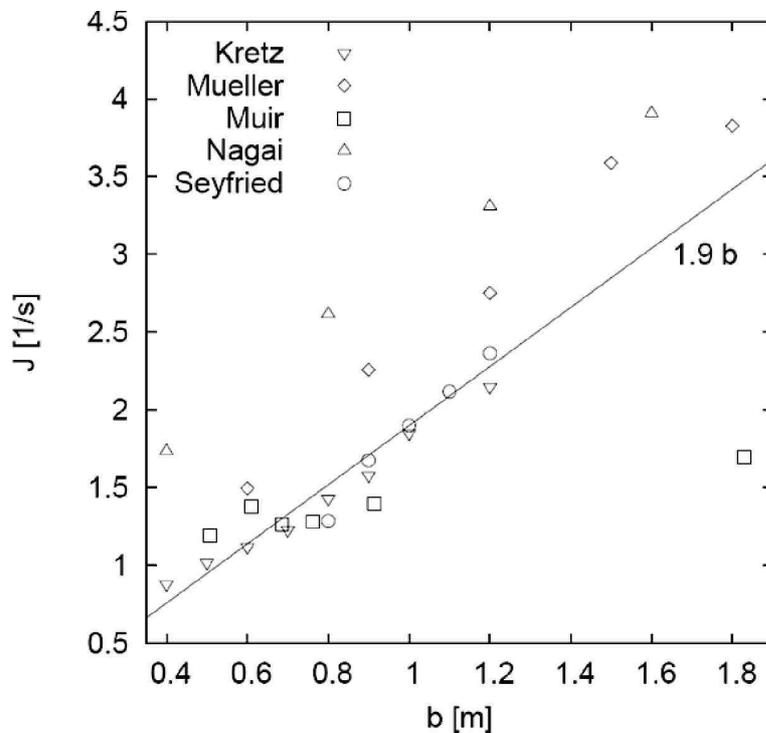

**Figure 2:** Flow in dependence of bottleneck width. Figure taken from [13].

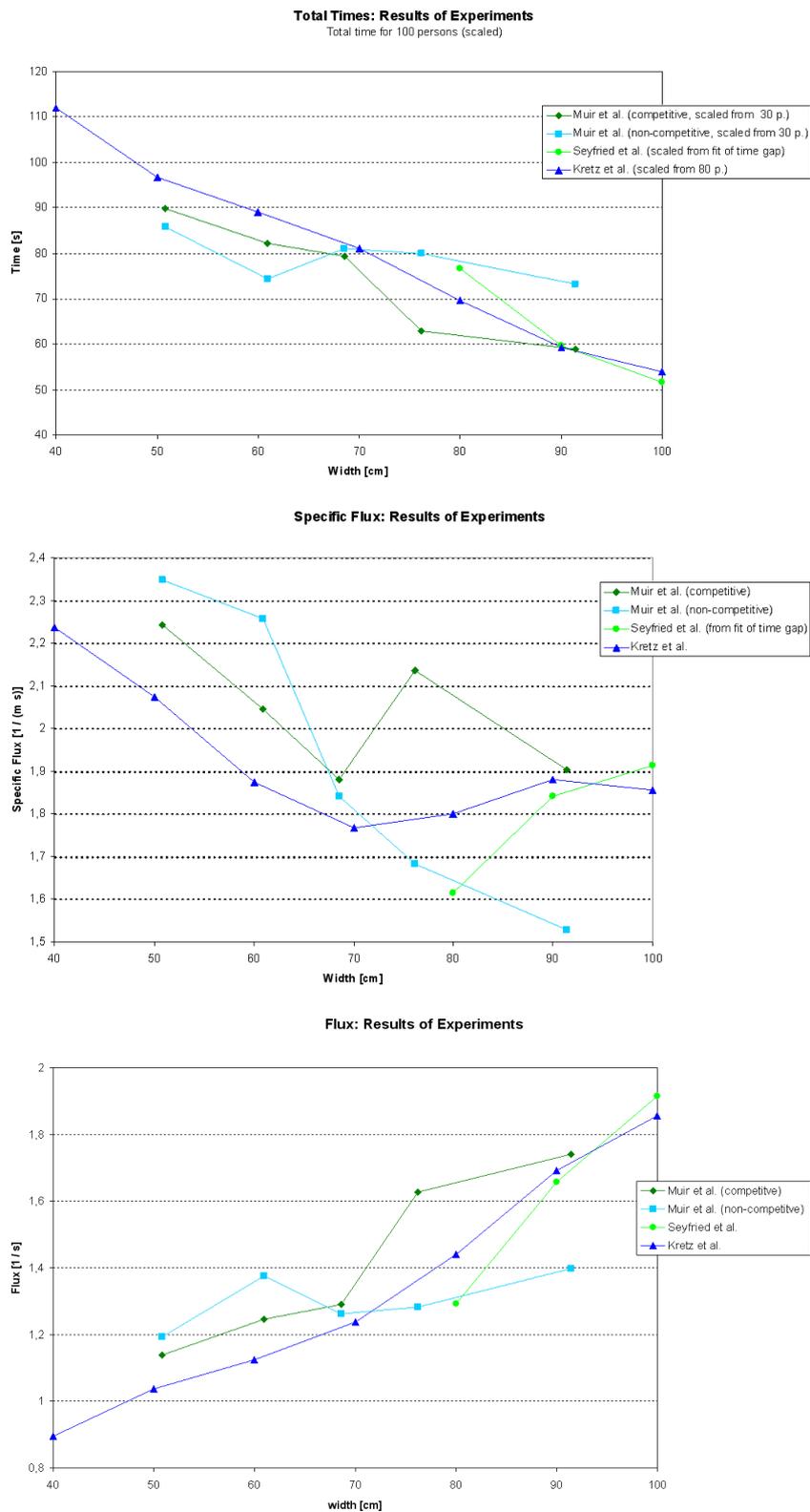

**Figure 3:** Total time, specific flux, and flux from various experiments (Muir et al. [5], Seyfried et al. [7], and Kretz et al. [6]). For the total time the results have been scaled such as if 100 persons had participated in the experiment.

**Interpretation of the Experimental Results**

The results of [5] and [6] deviate, both from the expectation of a constant specific flow and also from the expectation of a flow step function. They show a local minimum either in the function of the specific flow, which is less pronounced in [6] but clearly visible in the competitive variant of [5], or in the function for the flow (non-competitive in [5]), which is equivalent to a local maximum in the function for the total time. For a linear dependence between flow and bottleneck width, the specific flow should be constant, for a step function of the flow, the minimum in the function of the specific flow should be more pronounced: the specific flow should increase back to the maximum value at very small widths. In addition the minimum is expected at the width where two persons can pass the bottleneck shoulder-to-shoulder. Especially this last argument almost excludes the possibility of a linear dependence modulated by a small step effect as was assumed in [2] (compare figure 1).

So, the reason for the experimental findings must be something different. In [5] a psychological respectively perceptive phenomenon is stated:

> Contrary to expectation, more of these serious bottlenecks occurred in the 24-in. Configuration tests than in the 20-in. tests, even though the mean evacuation time for the former condition was less than that of the latter. [...] the 20-in. gap may have been perceived as too narrow to accommodate more than a single person passing through at any one time; and, consequently, escaping volunteers may have held back [...] .

This describes exactly, what has also been observed in [6] (compare figure 4) with the difference that the effect described in the quote was most pronounced at a width of 70 cm (not 60). Therefore, although the results of the single experimentation runs scatter significantly, one can be confident that the phenomenon exists and that it is responsible for the non-trivial function for time, and (specific) flow in dependence of the bottleneck width. The cause why the minimum for the specific flux in the competitive experiment of [5] is more pronounced than in [6] could be caused by the stronger motivation in [5]. This, however, would raise the question why then there is a minimum for the flux in the non-competitive variant of [5], which could be interpreted as most pronounced occurrence of the phenomenon. Another possible cause simply could be the scattering of results. Only further experiments can decide about that.

In a nutshell: what first might appear as a rudimentary step function in the function of the flow as shown in figure 3, namely a reduced increase between 50 and 70 cm and then again a strong increase between 70 and 80 or 90 cm width in fact is not caused by people fitting not and then fitting simultaneously into the bottleneck, but by two people not knowing if they fit into the bottleneck or not.

These observations demonstrate the richness of phenomena which one can use to calibrate and ever improve simulations of pedestrian traffic. On the other hand this makes it seem almost impossible to catch reality on such a detailed level in a fast computing simulation.

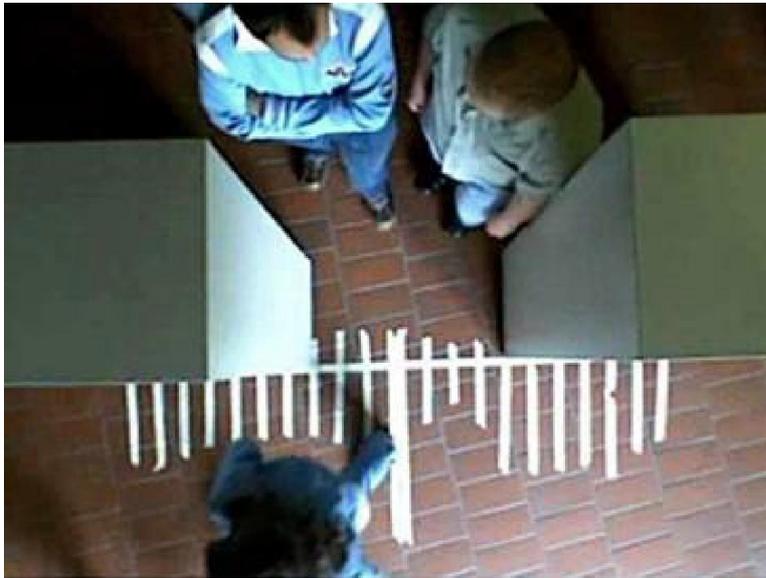

**Figure 4:** These two participants at first simultaneously approached to the 70 cm wide bottleneck to then hesitate when they realized they cannot pass at once. They first negotiated a passing order.

# BOTTLENECK SIMULATIONS WITH THE SOCIAL FORCE MODEL WITHIN VISSIM

### The Social Force Model within Vissim

PTV AG and the group of Dirk Helbing at ETH Zürich collaborate to integrate the Social Force Model into PTV s Vissim (see figure 5). Therefore the original Social Force Model [8, 9, 14 16] was adapted [17] to fit the needs of real life simulations. This affects the applicability to a vast amount of different scenarios and situations, population statistics, but also wholly new applications like the simulation of the interaction of pedestrians and vehicles.

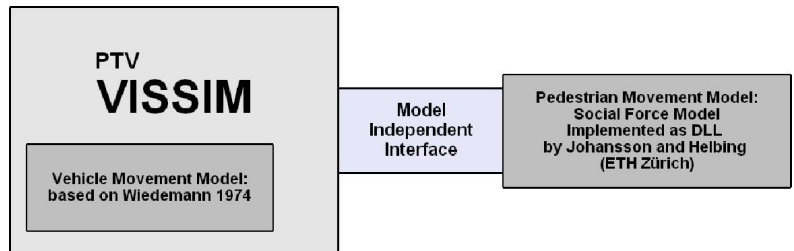

**Figure 5:** The Social Force Model and Vissim.

### Scenario

Figure 6 shows the simulation scenario. Six times 100 pedestrians were simulated to walk through six bottlenecks of various widths (40, 50, 60, 70, 80, and 100 cm). The scenario resembles especially the situation o the experiments in [6] in terms of initial density and initial distance between first pedestrians and bottleneck. The simulation was repeated ten times for each of the eight parameter sets which were investigated and the time when the last pedestrian of each of the groups passed the bottleneck was measured.

**Figure 6:** On each of the six blue areas 100 pedestrians were created which then moved through the bottlenecks of various widths toward the exit area (green) at the bottom edge.

### Simulation Parameters

The following parameters were used as parameter set zero (P0):

| Parameter | Value |
|---|---|
| radius of pedestrians | 15 cm |
| $A_{social}$ | 0.5 m/s² |
| $B_{social}$ | 2.8 m |
| $B_{physical,border}$ | 100 1/s² |
| $A_{social\ Isotropie}$ | 25 m/s² |
| $B_{social\ Isotropie}$ | 0.2 m |
| τ | 0.4 s |
| friction force | 0 |
| side preference | *right* |
| velocity dependence | 2 *s* |
| λ | 0.1 |
| longitudinal scale | 0.25 |
| consider at maximum *n* pedestrians | n = 5 |

Beginning with these parameters for the other seven parameter sets the following changes were made:

| Parameter Set | **Changed** | **To This Value** |
|---|---|---|
| *Pl* | $A_{social,\ isotropic}$ | 10 m/s² |
| *P2* | $A_{social,\ isotropic}$ | 100 m/s² |
| *P3* | $B_{social,\ isotropic}$ | 0.05 m |
| *PA* | $B_{social,\ isotropic}$ | 0.3 m |
| *P5* | *n* | 15 |
| *P6* | $A_{social}$ | 0.1 m/s² |
| *P7* | $A_{social}$ | 2.5 m/s² |

**Simulation Results**

Figure 7 shows the simulation results for eight respectively seven different sets of parameters. In all parameter sets except for P0 exactly one parameter was changed compared to P0.
With these results, the following facts are notable:
- The model is sensitive to these changes in the width of the bottleneck.
- Although different parameters were varied, the results almost entirely scaled with different constant factors.
- There is a minimum of the specific flux at 60 cm for five of the eight parameter sets. This could be due to
   o some artificial discretization effect,
   o a lane effect,
   o not a psychological effect, as this is this physical model.

The true reason has to be investigated in further studies.

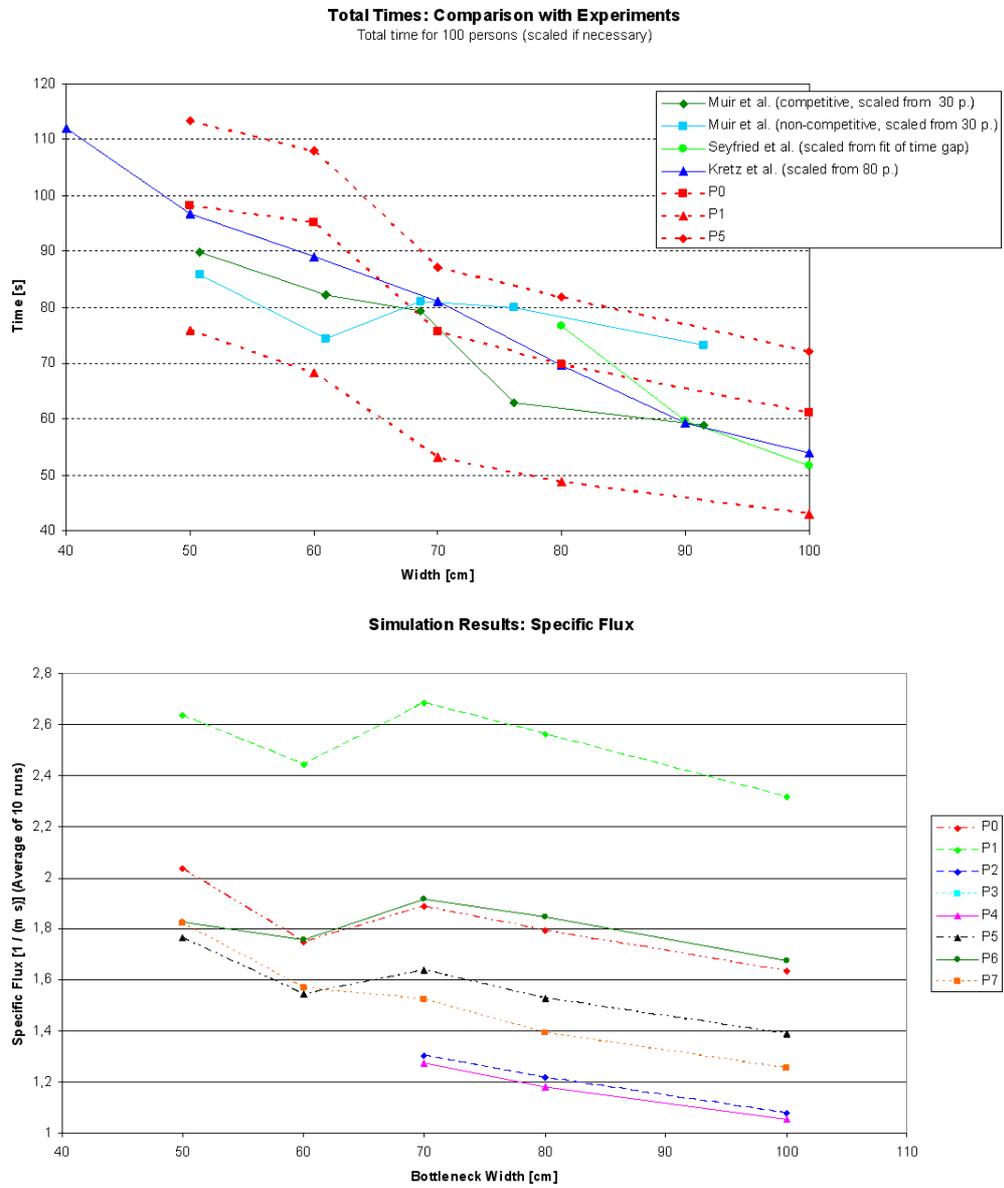

**Figure 7:** Total time and specific flux from simulations with different choices for the parameters. In the plot for the specific flux parameter set P3 was left out, as it resulted in far too large flows.

# COMPARISON OF EXPERIMENTS AND SIMULATION CONCLUSIONS

Figure 8 shows a comparison between experiments and simulation.

- In agreement between (some) experiments and simulation the specific flux shows a minimum. This is an astonishing success, considering the complex cause for the minimum in reality.
- The minimum was constantly found at smaller widths in the simulations than in the experiment(s). This could be because the radius of a pedestrian was set slightly too small.
- Apart from that, with the standard parameter set P0, the simulation gives results within the experimental result area (the range into which the experimental results fall).
- The model – just as reality – is sensitive to small changes of the bottleneck width.

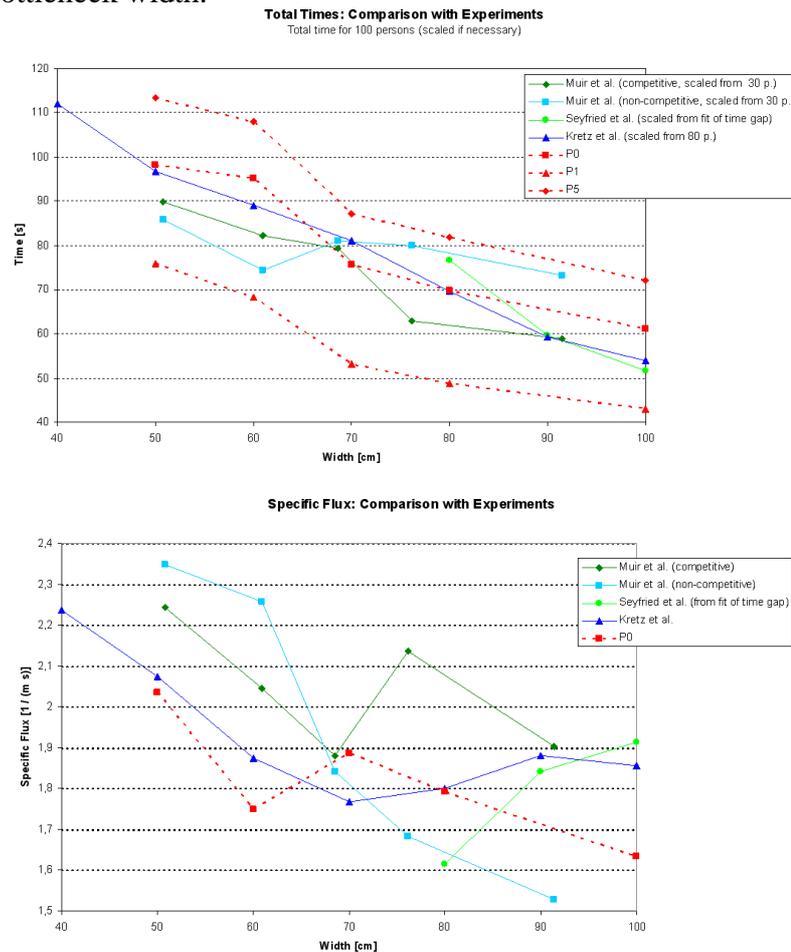

**Figure 8:** Comparison of the results from experiments and simulations.

# ACKNOWLEDGMENTS

We would like to thank D. Helbing, A. Johansson, C. Rogsch, and A. Seyfried for valuable support and discussion.